\documentstyle[preprint,aps]{revtex}
\begin{document}
\draft
\tighten
\title{ Application of Hypervirial Theorem As Criteria For Accuracy
    of Variational Trial Wave Function}
\author{Yi-Bing Ding}
\address{Graduate School, USTC at Beijing,
Academia Sinica, P.O.Box 3908, Beijing 100039, China; \\
and  Department of Physics, University of Milan, INFN,
20133 Milan, Italy}
\author{Xue-Qian Li}
\address{ CCAST(World Laboratory),P.O.Box 8730, Beijing 100080, China; \\
  and Department of Physics, Nankai University, Tianjin
300071, China}
\author{Peng-Nian Shen}
\address{ CCAST(World Laboratory),P.O.Box 8730, Beijing 100080, China;\\
Institute of High Energy Physics, Academia Sinica,
 P.O.Box 918(4), Beijing 100039, China;\\
and Center of Theoretical Nuclear Physics, National Laboratory
of Heavy Ion Accelerator, Lanzhou 730000, China}

\date{\today}

\maketitle

\begin{abstract}
It would be interesting to investigate the accuracy of the results obtained
in the variational method, because it is important for studying hadron spectra.
One can define some criteria to judge the accuracy, or the quality of the
trial function. We employ a simple potential form to check how accurate
the variational results obtained by a single-parameter trial function can be.
All the concerned problems, in particular, the relevant aspects on the
application of hypervirial theorem in variational method
for various potential forms, are discussed in
every detail. The qualitative conclusion of the work can be generalized to
much more complicated cases.
Our study suggests that the hypervirial relations can serve as a good and
practical criterion for judging the accuracy of any trial functions.
\end{abstract}
\pacs{ \bf PACS number(s): 03.65.Ge, 11.80.Fv, 12.39.Jh, 12.39.Pn}

\section{Introduction}

As well known, there are few potential forms whose bound state problems can
analytically be solved. Therefore, various approximate methods have been
used to do the job, in which, no doubt, numerically solving the Schr\"{o}dinger
equation is the most direct and powerful method, which in principle can meet
any desired accuracy. But on other side,
it cannot provide analytical expressions for further
discussions. Moreover, usually, as one numerically solves the Schr\"{o}dinger
equation, he can only keep a few terms in the zero-th order, while leaving
the others, such as the spin-dependent terms to be treated as a perturbation.
Indeed, the perturbative method is an approximate treatment
which has extensively been adopted because of its simplicity and convenience.
However, this method is limited by the
convergence of perturbative expansion. In fact, treating
wave functions is much more difficult than evaluating energy eigenvalues
\cite{MB}.

Another widely employed approach is the variational method.
Most desirable advantage of the method is that there exists an
analytical expression of the wave function for further discussions on physics.
In past years, lots of work paid their attention to seeking for
accurate energy eigenvalues but seldom discussed the wave function.
In many simple cases, the Gaussian-type
or the exponential-type trial wave functions (TWF)
with a single variational parameter are employed and they usually can
give an accurate value for the eigenenergy, but it is not true for the wave
function in general.

In a recent work \cite{DLS}, we investigated this problem in detail with
various potential forms which are used frequently in literatures to
study heavy quarkonia
and put more emphases on estimating the wave function at origin (WFO)
by using a TWF with single and several variational parameters.
We hope that it would shed some light on how to obtain
sufficiently accurate WFO in terms of the variational method.
We find that the accuracy of WFO seriously depends on
the choice of the forms of
TWF. In general, although there is no  universal rule to
determine the form and the number of the parameters of the TWF, one can
always find out a relatively simple and more reliable TWF with the
least number of parameters.  Our result showed an
economic way to realize it.

Recently, Lucha and Sch\"{o}berl have made a renewed investigation
\cite{LS} to qualitatively and quantitatively evaluate the accuracy, namely
how close to the exact eigenstates the results obtained in variational
method can be.  They embark upon a systematic study of the accuracy
of the variationally determined eigenstates of a Hamiltonian $H$ and
suitable measures to
judge their quality and apply their principles to the
prototype of all (semi-) relativistic bound-state equations, the ``spinless
Salpeter equation''. In their work,  the
harmonic-oscillator potential is employed as an example.
Their work is very instructive and stimulates
interest in studying the accuracy of the trial wave function for the
variational method. It is the goal of this work
to further study the accuracy of the
TWF with single variational parameter and the criteria to judge the accuracy.

In this work, we firstly extend the criteria used by the authors of
Ref.\cite{LS} and apply
them to treat a simple case of the non-relativistic
Schr\"{o}dinger equation where only  the linear potential exists
and give some further discussions on the accuracy of the TWF with
single parameter. The motivation to choose the linear potential is three-fold.
The first is its significance in the particle physics.
The next is that if only the linear potential exists, there are
exact solutions for nS states, which are the well-known Airy function Ai(z).
Thus, we
can conveniently  use them as a basis for comparing the results obtained with
the TWF and the exact
one  to quantify how close to each other the approximate solutions in the
variational method and exact eigenstates can be. The last one is that
we find that  the linear potential
possesses some interesting features which are helpful for a thorough
investigation of the criteria which can be generalized to other cases.

After the quantitative discussion to all the criteria for the ground states
and the lowest radial excitation state of the linear potential, we study in detail
the application of hypervirial relations as a most powerful and practical
criterion to judge the accuracy of the variational trial wave function. We
shall prove that when the variational parameter $a$ take its optimal value,
the special form of the TWF
$$R_{twf}^{(n)}(r)=N_n r^n e^{-a~ r^b}$$
satisfies the general virial theorem which is so called the first order
hypervirial relation. This is
independent of both the concrete form of the central potential and the specific
value of $b$. In this case we have to use higher order hypervirial relations
as a criterion to judge the quality of the TWF. To further testify the
applications of the hypervirial theorem, we calculate the relevant quantities
corresponding to several commonly adopted potential forms such as the Cornell,
logarithmic, Martin's and power-like ones.

The paper is organized as following. After the introduction,
we provide several criteria for judging the quality of the TWF, where the
linear potential is taken as an example to serve as the basis for discussion.
Indeed applications of all of them need to priori know the exact solutions for a
comparison, studying them is to find support to our main goal, i.e. to apply
the hypervirial relations as criteria. In Sec.III, we discuss applications of the
hypervirial relations which can serve as practical and useful criteria for the quality
of the TWF. In the last section, discussion and conclusion are made.

\vspace{1cm}

\section{Several exact-solution-dependent criteria for TWF}

\vspace{0.5cm}

What we concern is how close to the accurate solution
the obtained approximate wave function can be when the approximate
energy $E_{min}$ of Eq.(\ref {eq:E0} in the
Appendix B) is satisfactorily
close to the accurate energy eigenvalue, or in other words,  how seriously
the approximate $R_{twf}$ deviate from the real one and how
the nature and extent of these errors affect the calculated values.
There were intensive discussions on the
problem many years ago \cite{FW}. Recently Lucha and Sch\"{o}berl \cite{LS}
reviewed and developed these discussions and
introduced several measures to judge the quality of an
approximate wave function in the variational method, whose applications to our
case are described in the following.

(a) The distance between the obtained approximate value of the 
energy $E_{min}$ and
exact eigenvalue $E$. It can be represented by using the relative error
$\varepsilon $ of $E_{min}$ and $E$,

\begin{equation}
\varepsilon\equiv\frac{E_{min}-E}{E}\    .  \label{eq:REE}
\end{equation}

(b) The overlap integral $S$ of the trial state $|R_{twf}\rangle$ and
the eigenstate $|R_{eig}\rangle$.

If both $|R_{twf}\rangle$ and $|R_{eig}\rangle$ are
normalized, i.e.
$$\langle  R_{twf}|R_{twf}\rangle=\langle  R_{eig}|R_{eig}\rangle= 1,$$
one can define
\begin{eqnarray}
S=\langle R_{twf}|R_{eig}\rangle.
\end{eqnarray}
If the $R_{twf}$ is just the exact eigen-state $R_{eig}$, the overlap
$S$  is equal to unity. In general, we have
$$0\le|S|\le 1. $$
As a quantitative
criterion, according to the suggestion of authors of \cite{LS}, we take the
deviation of the squared modules of the overlap $S$,
from unity, $\sigma$ as
\begin{equation}
\sigma\equiv1-|S|^2. \label{eq:SIG}
\end{equation}

(c) The normalized maximum difference of variational function $R_{twf}(r)$ and
the eigenstate $R_{eig}(r)$, i.e., the maximum  relative
error,
\begin{eqnarray}
\omega\equiv\frac{\displaystyle\max_{\bf r}[R_{twf}(r)-R_{eig}(r)]}
{\displaystyle\max_{\bf r}R_{eig}(r)}. \label{eq:REW}
\end{eqnarray}

Besides the above criteria, we introduce the following
measures which may reveal the physical properties of the TWF.

(d) The squared wave function at origin (WFO) $R^2(0)$.

As is noticed in our earlier work \cite{DLS}, this quantity can be obtained
by using two methods. The first approximate WFO,
$R^2(0)$~{\bf I},
is directly calculated from the normalized radial wave function by
setting $r=0$.
And in the second approach, $R(0)^2$ {\bf II} can be obtained
from the well-known equation \cite{QR}
\begin{eqnarray}
R^2(0)=2\mu \langle~R|\frac{dV}{dr}~|R~\rangle.   \label {eq:R02}
\end{eqnarray}
If $R(r)$ is the real solution of the Schr\"{o}dinger equation with the
given potential, the resultant $R(0)^2$ {\bf I} and $R(0)^2$ {\bf II}
should exactly be the same.
However, if we use $R_{twf}(r)$ to calculate these two quantities, 
the results definitely deviate from each other.
Our numerical results  \cite{DLS} showed
that the later one can generally reach very high accuracy. 

In the present case of the linear potential $V(r)=r$ with $2\mu=1 GeV$,
it is very easy to prove that the resultant $R(0)^2$ {\bf II} is
\begin{eqnarray}
R^2(0)=\langle R|R\rangle.   \label {eq:R03}
\end{eqnarray}
Thus, for the any $R(r)$, no matter it is an accurate eigen-wavefunction
$R_{eig}(r)$ or any  TWF $R_{twf}(r)$, so long as it satisfies the
normalization
condition, the $R(0)^2$ obtained by using such method must be identical
to unity.
Definitely, this is a special feature
possessed by the $S$-wave bound state for the linear potential. Therefore, we
only have the first  type of the WFO $R(0)^2 ${ \bf I}
can be used in our case. In practice,
what we take
as a quantitative criterion is the relative deviation
\begin{equation}
\delta r_{02} \equiv \frac{R_{twf}^2(0)-R_{eig}^2(0)}{R_{eig}^2(0)}.
\end{equation}
However, in the general case, $dV/dr\neq 1$, then both $R^2(0)${\bf I} and
$R^2(0)${\bf II} can be used to obtain $\delta_{r_02}$.

(e) The average value of the operators $r^2$ and $r^{-2}$.

In order to show the effect of the approximate wave function on some
particular properties, we calculate average values $\langle r^2 \rangle$
and $\langle r^{-2} \rangle$. In fact, the weighted overlap
integrals \cite{FW}
depend principally on regions of the configuration space. The accuracy of
$\langle r^2 \rangle$ depends
mainly on the accuracy of TWF in the outer regions
and that of $\langle r^{-2} \rangle$  depends more sensitively on the
region near the
origin. In practice, what we select as the criteria are the relative errors
of the approximate value with respect to the exact one
\begin{eqnarray}
\delta r_{p2} \equiv \frac{\langle R_{twf}| r^2|R_{twf}\rangle - \langle R_{eig}|r^2
|R_{eig}\rangle}{\langle R_{eig}|r^2|R_{eig}\rangle}
\end{eqnarray}
and
\begin{eqnarray}
\delta r_{n2} \equiv \frac{\langle R_{twf}| r^{-2}|R_{twf}\rangle -
\langle R_{eig}|r^{-2}|R_{eig}\rangle}{\langle R_{eig}|r^{-2}|R_{eig}\rangle}.
\end{eqnarray}

 All the numerical results corresponding to these criteria for
the ground state and low excitations of the linear potential,
including the values of the optimal variational parameter $a_0$,  the
minimum values of the energy $E_{min}$,
are listed in  Table 1.

The numerical results in Table 1 confirm that all the criteria can be 
effectively used for judging the quality of the TWF.\\

\section{The hypervirial relations as most practical and
           powerful criteria}

\vspace{0.1cm}

\subsection{The hypervirial theorem and hypervirial relations}

The virial theorem has been widely used in both classical and
quantum mechanics. In quantum mechanics, it can be formulated as
$$2<T>=<\vec r\cdot\nabla V(r)>,$$
where $T$ is the kinetic energy operator and
the expectation value is taken for eigen-wavefunctions of the Hamiltonian $H$.
It holds for any central potential $V(r)$.

It is easy to prove that an equivalent statement of the virial theorem is 
that the expectation value of the commutator $[W,H]$ on 
the energy eigenstates $|R_{eig}\rangle$, vanishes:
\begin{equation}
\langle R_{eig}|[W,H]| R_{eig}\rangle=0\ .\label{eq:VC0}
\end{equation}
where
\begin{equation}
W=\frac{1}{2} ({\bf r \cdot p}+ {\bf p \cdot r}), \label{eq:Dia}
\end{equation}
is called as the generator of dilatations and $H$ is the Hamiltonian.

Hirschfelder generalized the theorem to the so called hypervirial
theorem \cite{JH}. It states that the expectation value
of the commutator $[G,H]$, where $G$ stands for any time-independent 
linear operator, on the energy eigenstate $|R_{eig}\rangle$ is equal 
to zero:
\begin{equation}
\langle R_{eig}|[G,H]| R_{eig}\rangle=0\ .\label{eq:VC}
\end{equation}
Obviously our $W$ is such an operator.

However, for a trial state $|R_{twf}\rangle$,
in general, the expectation value

\noindent $\langle R_{twf}|[G,H]| R_{twf}\rangle$
is not equal to zero. We can expect that the closer to the real
solution the
approximate trial function is, the closer to zero the expectation value
would be. Hence, a set of operators $G$ generates a whole set of the
"hypervirial relations", which will be defined in Eq.(\ref {eq:GH1}),
 and each of them may
serve as a measure of the quality of a given trial function $|R_{twf}\rangle$
by evaluating the expectation value
$\langle R_{twf}|[G,H]|R_{twf}\rangle$.

In the present work,  the hypervirial operator can be chosen as \cite{JK,LG}
\begin{eqnarray}
G=f(r)p_r
\end{eqnarray}
where $f(r)$ is a function of $r$ and $p_r$ is the radial momentum operator
conjugate to $r$, whose explicit expression is
$p_r=-i(d/dr + 1/r)$. The corresponding Hamiltonian (with $2 \mu=1$ GeV)
reads
\begin{eqnarray}
H=(p_r)^2+\frac{l(l+1)}{r^2}+V(r).   \label {eq:HD}
\end{eqnarray}
Considering
\begin{eqnarray}
[G,H] &\equiv& [f(r)p_r,H] \nonumber \\
      &=&2 i \frac{d f(r)}{dr} p_r^2 + i \frac{d^2 f(r)}{dr^2} p_r
         +2 i f(r) \frac{l(l+1)}{r^3}- i f(r) \frac{dV}{dr},
\end{eqnarray}
\begin{eqnarray}
p_r^2=H-\frac{l(l+1)}{r^2}-V
\end{eqnarray}
and
\begin{eqnarray}
\frac{d^2f(r)}{dr^2}p_r =\frac{i}{2}\frac{d^3f(r)}{dr^3}-
      [\frac{df(r)}{dr},H],
\end{eqnarray}
we can get the commutator
\begin{eqnarray}
[G,H]&=& 2 i \frac{d f(r)}{dr}(H-V)+ 2 i l(l+1) (\frac{f(r)}{r^3}-
       \frac{1}{r^2}\frac{df(r)}{dr})+\frac{i}{2}\frac{d^3f(r)}{dr^3}
       \nonumber \\
     &-&i f(r) \frac{dV}{dr}-\frac{i}{2}[\frac{df(r)}{dr},H]. \label{eq:GH}
\end{eqnarray}

The hypervirial theorem demands the expectation values of both $[G,H]$ and
$[\frac{df(r)}{dr},H]$ for the eigenstate of $H$ to be zero, thus from
Eq. (\ref{eq:GH}) we can obtain an algebraic equation
\begin{eqnarray}
2E\langle \frac{df(r)}{dr} \rangle
    &=& 2 \langle V\frac{df(r)}{dr} \rangle
       + \langle f(r) \frac{dV}{dr} \rangle -
       2 l(l+1) (\langle \frac{f(r)}{r^3} \rangle-
       \langle \frac{1}{r^2}\frac{df(r)}{dr}\rangle) \nonumber \\
      & -& \frac{i}{2}\langle \frac{d^3~f(r)}{dr} \rangle. \label{eq:GH1}
\end{eqnarray}
where $\langle \cdots \rangle$ stands for the expectation value
over the eigenstate.
These algebraic equations (\ref{eq:GH1})
are defined as the hypervirial relations.

Taking
$$ f(r)=r^N  \ \ \ \ \ (N=1,2,3, \cdots), $$
Eqs. (\ref{eq:GH}) and (\ref{eq:GH1}) can respectively be re-formed as
\begin{eqnarray}
[G_N,H] &\equiv & [r^N~p_r,H] \nonumber \\
&=&2 i Nr^{N-1}(H-V)-ir^N \frac{dV}{dr}-2 i l(l+1) (N-1)r^{N-3}
 + \frac{i}{2}N(N-1)(N-2)r^{N-3}\nonumber \\
 &-& \frac{i}{2}[N r^{N-1},H]   \label {eq:WH}
\end{eqnarray}
and
\begin{eqnarray}
2EN\langle r^{N-1} \rangle &=& 2N \langle r^{N-1}V \rangle +
   \langle r^N(DV) \rangle+2 l(l+1)(N-1) \langle r^{N-3} \rangle \nonumber
   \\
  &-& \frac{1}{2}N(N-1)(N-2) \langle r^{N-3} \rangle,   \label {eq:RR1}
\end{eqnarray}
where $G_N=r^N p_r$ is the adopted hypervirial operators 
and Eq.(\ref {eq:RR1})
is an important recurrence relation which has extensive applications.

For the $nS$ state in the linear potential case, we can obtain,
from Eq.(\ref{eq:GH1}), a set of very useful
recurrence relations
\begin{eqnarray}
2EN\langle r^{N-1} \rangle = (2N+1) \langle r^N \rangle -
  \frac{1}{2}N(N-1)(N-2) \langle r^{N-3} \rangle,   \label {eq:RR2}
\end{eqnarray}
which can be named the $N$-th order hypervirial relation.

When $N=1$, we have the first order hypervirial relation
\begin{eqnarray}
2E= 3 \langle r \rangle,   \label {eq:RR3}
\end{eqnarray}
which is just the general virial theorem in the linear potential case.
Taking $N=2$ and $N=3$, we can obtain the second and the third order
hypervirial relation, which  respectively are
\begin{eqnarray}
4E\langle r \rangle=5 \langle r^2 \rangle   \label {eq:RR4}
\end{eqnarray}
and
\begin{eqnarray}
6E\langle r \rangle=7 \langle r^3 \rangle-3.
\end{eqnarray}
In the above equations, the average is taken over the eigenstate of $H$.

\subsection{Deviation from the hypervirial relations
   as criteria for accuracy of TWF}

As mentioned in the last subsection, for a properly chosen 
trial function $R_{twf}(r)$, these hypervirial relations
should merely approximately hold. According to the
1-st, 2-nd and 3-rd hypervirial relations,
we can respectively define deviations:
\begin{eqnarray}
\nu_1\equiv 2E-3 \langle r \rangle,   \label {eq:NU1}
\end{eqnarray}
\begin{eqnarray}
\nu_2\equiv 4E \langle r \rangle-5 \langle r^2 \rangle, \label {eq:NU2}
\end{eqnarray}
and
\begin{eqnarray}
\nu_3\equiv -6E \langle r^{2} \rangle + 7 \langle r^3 \rangle -3, 
\label {eq:NU3}
\end{eqnarray}
where the average value of $r^N$ can be analytically calculated
in terms of our TWF\footnote{It is noted that in this subsection, the $<>$
stands for an average with respect to the trial wavefunction TWF}.
For the ground state, by using Eq. (\ref{eq:TWF})
with the optimal variational
parameter $a_0$, one has 
\begin{eqnarray}
\langle r^N \rangle=\frac{\Gamma(\frac{3+N}{b})}
                         {\Gamma(\frac{3}{b})}
    \Bigl( \frac{(1+b)\Gamma(\frac{1}{b})}
               {2\Gamma(\frac{4}{b})}\Bigr)^{\frac{N}{3}}.     \label {eq:AR}
\end{eqnarray}
In terms of Eq.(\ref{eq:AR}) and $E_{min}$,
we can easily obtain the numerical values of $\nu_1$, $\nu_2$
and $\nu_3$. In order
to evaluate hypervirial relations for judging
the quality of TWF, we further
introduce relative errors $\delta \nu_i$ (i=1,2,3...) instead of $\nu_i$ as
\begin{eqnarray}
         \delta \nu_1  \equiv  \frac{|\nu_1|}{3 \langle r \rangle}, \ \ \ \
         \delta \nu_2  \equiv \frac{|\nu_2|}{5 \langle r^2 \rangle},\ \ \ \
         \delta \nu_3  \equiv \frac{|\nu_3|}{7 \langle r^3 \rangle}.
\label {eq:DM}
\end{eqnarray}
Obviously, these expressions (\ref{eq:DM}) of $\delta\nu_i$ are only valid for
the linear potential, for other potential forms, the denominators in
(\ref{eq:DM}) should be replaced.

The results of $\nu_1, \delta \nu_1 , \nu_2, \delta \nu_2 ,
 \nu_3 $ and $\delta \nu_3$  are listed
in Table 1, respectively.
In the next section, we will give more discussions on
the implications of these results.

In this work, we specially choose the most frequently used hypervirial
operators $G_N=r^N p_r$. When $N=1$, $G_1=r p_r$ has a clearly physical 
meaning. Although it itself is not a Hermitian operator, it only
differs from its symmetrized and  Hermitian form $W$ (see Eq.(\ref{eq:Dia}))
by a constant, i.e.,
\begin{equation}
W=G_1- \frac{i}{2}.
\end{equation}
As a consequence, $$ [W,H]=[G_1,H], $$ and one obtains the same 
first order hypervirial relation.

As well known \cite{WL}, the dilation generator $W$ generates the
scaling transformation or dilation of the phase
space, which is an important transformation not only to quantum mechanics
but also to classical mechanics. Its classical equivalent 
${\bf r} \cdot {\bf p}$ has
been used to derive a virial theorem in the classical mechanics.

Using the hypervirial relation of the quantum mechanics derived in terms of
the operators  $G_N$, one can derive many useful recurrence relations 
between $\langle r^N \rangle$ and $E$. These relations can be used
to gain approximate eigen-spectra and the
expectation values of observables without solving the wave
functions of bound states as well as to  calculate the phase shifts and the
amplitudes of the scattering processes more effectively than in the 
perturbative method. In this investigation, what we concern is 
another important application noticed by Lucha and Sch\"{o}berl 
\cite{LS}. We will show that the
hypervirial relations could be sensitive criteria for the accuracy of
the solution by using the variational method. Of course, if we choose a more
complicated form, we cannot have such compact expressions and the results
of $\nu_i$ and $\delta\nu_i$ would be presented only numerically. Even so,
the principle and the general conclusion hold.

In the last section, we give many criteria which can be used to judge
the accuracy of TWF. It can be seen from Table 1 that all these criteria
give a consistent
result, by which we can select a most optimal TWF corresponding to
$b=\frac{7}{4}$ for 1S state of the linear potential from four possible
value of $b$. However,  the practical significance
of these criteria to judge the accuracy of the TWF are limited by
their dependence on the exact solution. It is clear that if one can get
the exact solution, it, in general, would
be unnecessary to find the approximate solution. But sometimes we want
to use the variational method to find a simple analytical form
of the approximate wave function with single parameter or a few parameters
to replace a complicate exact solution or a pure numerical solution, and then
these criteria are certainly very helpful.
However, for those problems that we do not need an exact solution or
it is very difficult to obtain the exact
solution, these exact-solution-dependent quantities have no practical usage.
In the case, as criteria
the deviations from the hypervirial relations are the unique choice
because only they do not require any knowledge on the exact eigen state
and consequently can be applied to all the cases.

\vspace{0.5cm}

\centerline{ Table 1 }

\vspace{0.3cm}

{\footnotesize
The values of various criterion quantities for the single-parameter
variational
solutions of the $1S$, $2S$ and $3S$ states in the linear potential
case. The exact eigenvalues of energy are $E_{1S}=2.33810 GeV$,
$E_{2S}=4.08794 GeV$ and $E_{3S}=5.52055 GeV$, respectively.}

\vspace{1cm}

\begin{center}
\begin{tabular}{|c|c|c|c|c|c|c|}
\hline
 Quantity  & \multicolumn{4}{|c|}{$1S$ state}& $2S$ state& $3S$ state\\
\hline
b&1 & $\frac{3}{2}$ &  $\frac{7}{4}$ &  2  &$\frac{7}{4}$&$\frac{7}{4}$ \\
\hline
$a_0$&0.908560 & 0.471405&0.348957& 0.260531& 0.348957&0.348957\\
\hline
$E_{min}$ (GeV) & 2.47645&2.34723 & 2.33825 & 2.34478&4.09917&5.52874 \\
\hline
$\varepsilon$&0.05917&0.00390&  0.00006& 0.00285&0.00275&0.00143\\
\hline
$\sigma$& 0.032731 &0.00207&0.00003&0.00180&0.00634&0.22584\\
\hline
$\omega$& 0.14303&0.04353& -0.00436& 0.35934&-0.0808&0.63515\\
\hline
$\delta r_{02}$ & 2.00000&0.33333&0.03646 & -0.15117&-0.12956&-0.73780\\
\hline
$\delta r_{p2}$&0.24648&0.03070&-0.00086&-0.01264&-0.09030&-0.16022\\
\hline
$\delta r_{n2}$&0.46777&0.11295&-0.00790&-0.07351&0.01229&-0.30906\\
\hline
$\nu_1$ &0       &0       &0     &0     &0.13701&0.23532\\
\hline
$\delta\nu_1$ &  0 &    0 &  0 &   0 &  0.05278 &  0.06820 \\
\hline
$\nu_2$&-1.81712& -0.33349& 0.01439&0.26760 &2.02267&8.04999\\
\hline
$\delta\nu_2$ &  0.10000 & 0.02220 & 0.00099&  0.01859 & 0.04989&  0.11795\\
\hline
$\nu_3$&13.000 & 1.9280 & -0.13021 & -1.5000 & -9.7304 & -53.865\\
\hline
$\delta\nu_3$ &  0.18571 & 0.04000 & 0.00298 &  0.03571 & 0.50498 & 0.13402\\
\hline
\end{tabular}
\end{center}

\vspace{0.5cm}

\subsection{Restriction on applying hypervirial relations}

As noticed, there are some cases in which the expectation value
$\langle R_{twf}|[G,H]| R_{twf}\rangle$ vanishes accidentally. We list
following possible cases:

(i) TWF is an eigenstate of $G$,

(ii) TWF is an eigenstate of the commutator  $[G,H]$ with eigenvalue $0$,

(iii) $G|R_{twf}\rangle$ is orthogonal to $H|R_{twf}\rangle$,

(iv) $[G,H]|R_{twf} \rangle$ is orthogonal to $|R_{twf}\rangle$.

Evidently,  in these cases the corresponding hypervirial relation
cannot be used as a
criterion.  Table 1 indicates that we just confront this case for
the $1S$ state in the linear potential. The optimal TWF
(\ref{eq:TWF}) gives
$\nu_1\equiv 0$ and the result is independent of the value
of $b$.

In fact, since we choose $G_1=r p_r$ as the first order hypervirial operator
which is just the
generator of the dilation $W$ and  has no Hilbert-space eigenvector
{\cite{LS}, with N=1 and Eq.(\ref {eq:WH}), we have
\begin{eqnarray}
[rp_r,H]=3r-2H.   \label {eq:RH}
\end{eqnarray}
One can straightforwardly prove that for the optimal TWF of the ground state,
$R_{twf}(r)$ in the form of Eq.  (\ref{eq:TWF}),
$[rD,H]|R_{twf}\rangle$ is orthogonal to the state
$|R_{twf}\rangle$. Thus, this is the case (iv)  and
the $1$-st order hypervirial relation cannot be used to judge the quality
of the concerned TWF. Definitely, in this case,
we can use $\nu_2$ which is not equal to zero and can sensitively
judge the accuracy of the concerned TWF.
To our knowledge, the authors of most literatures only discussed the possibility
for the case where the hypervirial relation fails to testify the given
trial function, but seldom gave concrete examples. The form of
TWF (\ref{eq:TWF}) presented by us is a convincing illustration.

We find that at this point we can give stronger arguments and generalize
it to extensive cases. Epstein and Hirschfelder have proved a
well-known lemma \cite{EH} which states that
if a variational trial function $\psi$ obeys
\begin{eqnarray}
\frac{\partial \psi}{\partial a}=i W \psi  \label {eq:PSI1}
\end{eqnarray}
where $a$ is the variational parameter and $W$ is a Hermitian operator,
then the optimal  $\psi$ satisfies the
hypervirial theorem for the Hermitian
operator $W$, i.e.
$$\langle \psi| [W,H] |\psi\rangle =~0.$$
The authors of ref. \cite{HC} generalized the lemma and claimed that
if Eq. (\ref {eq:PSI1}) is replaced by
\begin{eqnarray}
\frac{\partial \psi}{\partial a}= B i W \psi + i C \psi, \label{eq:PSI2}
\end{eqnarray}
the lemma is still correct. In Eq.(\ref{eq:PSI2}) $B$ and $C$ are the 
functions of $a$ , $B$ is real and $W$ is Hermitian. 
Evidently, Eq.(\ref{eq:PSI1}) is a special case of
Eq.(\ref{eq:PSI2}) with $B=1$ and $C=0$. This lemma and its
generalized form have extensively been used to select
approximate TWF's and to discuss the spectroscopy of
molecule and atom.

Later, as a sample of the application of the
generalized lemma, we will stay in the linear potential case, unless
it is specially specified.

With this generalized lemma, we can deduce some very useful consequences.

(a) For the $1S$ state, when TWF is in the form (\ref{eq:TWF}),
we have
\begin{eqnarray}
\frac{\partial R_{twf}(r)}{\partial a}
   = (\frac{3}{2 a b}-r^b) R_{twf}(r). \label {eq:PSI3}
\end{eqnarray}
As aforementioned, taking
\begin{eqnarray}
W=\frac{1}{2} ( rp_r+p_rr) =-i (r \frac{d}{dr}+\frac{3}{2})
\end{eqnarray}
as the hypervirial operator, we get
\begin{eqnarray}
i W R_{twf}(r)= ( \frac{3}{2}-a b r^b ) R_{twf}(r).  \label {eq:WR1}
\end{eqnarray}
Comparing (\ref {eq:WR1}) with (\ref {eq:PSI3}), we can obtain
\begin{eqnarray}
\frac{\partial R_{twf}(r)}{\partial a}
   = \frac{1}{ab} i W R_{twf}(r),   \label {eq:RAW}
\end{eqnarray}
which corresponds to $B=-\frac{1}{ab}$ and $C=0$ in Eq.(\ref {eq:PSI2}).
When the variational parameter $a$ takes its optimal value,  we must have
 $\nu_1=0$, which is independent of the value $b$.
This means that the optimal TWF
(\ref {eq:TWF}) for the $1S$ state satisfies 
the general virial theorem although it is not the exact wave function.
If the employed TWF is of this kind, we at least have to use
the second order hypervirial relation $\nu_2$.

(b) In fact, (a) can be generalized to the other cases, because the result
is independent of the concrete form of the central potential $V(r)$.
As long as we take Eq. (\ref {eq:TWF}) as TWF for the $1S$
state in any central potential case, the general virial
theorem must be satisfied, namely $\nu_1=0$. The general virial theorem
cannot be taken as a criterion
to judge the quality of TWF. Then, we have to use the deviations from
2-nd and 3-rd order
hypervirial relations $\nu_2$ and $\nu_3$ as the criteria.

To investigate the applications of the above consequence, we
evaluate $\nu_i'$s with the frequently used potential forms
including
(i)the Coulomb potential, (ii) the Cornell potential $V(r)=-\frac{1}{r}+r $,
(iii) the Martin potential $V(r)=r^{0.1}$,
(iv) the logarithmic potential $V(r)=log(r)$
and (v) the three dimensional isotropic oscillator potential
$ V(r)=r^2$, respectively. The expressions of $\nu_i$ and $\delta \nu_i$
( i=1,2,3) obtained from the general formula and the corresponding
numerical results for these potentials are shown in Table 2.

It can be seen from the table that for the cases , (ii) , (iii) and (iv),
when $b=\frac{3}{2}$, one obtains the best TWF. On the other hand, in the
case (i) (or (v)), the closer to 1 (or 2) the value of $b$ is, the smaller the
corresponding $\delta\nu_2$ and $\delta\nu_3$ are. Eventually, when $b=1$
(or $2$), the TWF is just the exact solution for the corresponding
potential
and all $\nu_i$ and $\delta \nu_i$ go to zero. Obviously, one can
use $\delta\nu_2$ and $\delta\nu_3$ as the criteria to judge whether TWF
is accurate enough.

\vspace{0.5cm}

\centerline{ Table 2 }
\vspace{0.3cm}
{\footnotesize
Variational results with the single-parameter TWF (\ref {eq:TWF})
in some  central potentials cases mentioned in the text.
}

\begin{center}
\begin{tabular}{|c|c|c|c|c|}
\hline
  $b$ & 1&   $\frac{3}{2}$  & $\frac{7}{4}$ &  2\\
\hline
\hline
\multicolumn{5}{|c|}{Coulomb potential $V(r)=-\frac{1}{r}$}\\
\hline
$a_0$&  0.50000 & 0.19159 &  0.11770 & 0.07074\\
\hline
$E$&  -0.25000 & -0.23555 & -0.22400 & -0.21221\\
\hline
$\nu_1\equiv -\langle 1/r \rangle -2E$&0& 0&0& 0\\
\hline
$\nu_2\equiv -3-4E \langle r \rangle $&  0  &  -0.31289 & -0.39463 & -0.45352\\
\hline
$\nu_3 \equiv -5 \langle r \rangle-6E\langle r^2 \rangle -3$
& 0  &  -3.1523 & -3.9158 & -4.5000\\
\hline
$\delta \nu_2\equiv \frac{|\nu_2|}{3}$& 0 &  0.10439 & 0.13154 & 0.15117\\
\hline
$\delta\nu_3 \equiv \frac{|\nu_3|}{5 \langle r \rangle }$
& 0  &   0.01543 & 0.01931 & 0.02122 \\
\hline
\hline
\multicolumn{5}{|c|}{Cornell potential $V(r)=-1/r+r$}\\
\hline
$a_0$&1.10939 & 0.63544 &  0.49016 &  0.37969 \\
\hline
$E$& 1.47345&1.39969 &  1.41824 & 1.45064 \\
\hline
$\nu_1\equiv 3 \langle r \rangle -\langle \frac{1}{r} \rangle-2 E$& 0 &0 & 0 & 0\\
\hline
$\nu_2 \equiv 5 \langle r^2 \rangle -4 E \langle r \rangle -3 $
&1.21878 &-0.08892 & -0.40439& -0.63714\\
\hline
$\nu_3 \equiv 7\langle r^3 \rangle -6E\langle r^2 \rangle - 5 \langle r \rangle-3$
& 7.1409 & -0.35635 & -1.8015 & -2.7949\\
\hline
$\delta \nu_2 \equiv \frac{|\nu_2|}{5 \langle r^2 \rangle}$
&0.10000 &0.00881&0.04094 &0.06514\\
\hline
$\delta\nu_3 \equiv \frac{|\nu_3|}{7 \langle r^3 \rangle}
$ &0.18571 & 0.01370 &  0.07375 & 0.11708\\
\hline
\hline
\multicolumn{5}{|c|}{Martin potential $V(r)=r^{0.1}$}\\
\hline
$a_0$ & 0.24300&0.06770& 0.03641 & 0.01965 \\
\hline
$E$ &1.24007&1.23576&1.23654&1.23811\\
\hline
$\nu_1 \equiv 2.1 \langle r^{0.1} \rangle -2 E$&0 & 0 &0 & 0\\
\hline
$\nu_2 \equiv 4.1  \langle r^{1.1} \rangle -4 E \langle r \rangle$
&0.26731 & -0.01636 &-0.08947 & -0.14441\\
\hline
$\nu_3 \equiv 6.1 \langle r^3 \rangle -6E \langle r^2 \rangle -3$ & 6.6550 & -0.41743 & -1.8517 & -2.8500 \\
\hline
$\delta \nu_2 \equiv \frac{|\nu_2|}{4.1 \langle r^{1.1} \rangle} $& 0.00865 &0.00058&0.00320& 0.00515\\
\hline
$\delta\nu_3 \equiv \frac{|\nu_3|}{6.1 \langle r^3 \rangle}$
&0.01717 & 0.00140 & 0.00645 & 0.01005 \\
\hline
\hline
\multicolumn{5}{|c|}{Logarithmic potential $V(r)=log(r)$}\\
\hline
$a_0$&0.70711&  0.33693& 0.23663& 0.16667\\
\hline
$E$&1.07521& 1.0450& 1.05310& 1.06755\\
\hline
$\nu_1 \equiv 2 \langle log(r) \rangle +1 -2 E$&0& 0&0& 0\\
\hline
$\nu_2 \equiv 4 \langle r~ log(r) \rangle -( 4 E-1) \langle r \rangle$
&0.70711 &-0.08362&-0.28942&-0.44455\\
\hline
$\nu_3 \equiv 6 \langle r^2 log r \rangle -(6E-1)\langle r^2 \rangle-3$
&6.0000 &  -0.67223 & -2.0412 & -3.0000 \\
\hline
$\delta \nu_2 \equiv \frac{|\nu_2|}{4 \langle r~log(r) \rangle}$
&0.09162&0.01362&0.04854&0.07474\\
\hline
$\delta\nu_3 \equiv \frac{|\nu_3|}{6 <r^2 log r>}$
& 0.14374 &  0.02480 &  0.08129 & 0.12334 \\
\hline
\hline
\multicolumn{5}{|c|}{3-D isotropic oscillator potential $V(r)=r^2$}\\
\hline
$a_0$&   1.3161 & 0.78082 &  0.52127 & 0.50000\\
\hline
$E$&  3.4641 & 3.0667 &   3.0136 &  3.0000\\
\hline
$\nu_1 \equiv \langle r^2 \rangle -E/2 $ &0& 0&0& 0\\
\hline
$\nu_2 \equiv 6 \langle r^3 \rangle -4 E \langle r \rangle$
&   3.9482 &   1.0499 & 0.43243  &   0 \\
\hline
$\nu_3 \equiv 8 \langle r^4 \rangle -6 E \langle r^2 \rangle -3 $
&  21.000  &  4,63819  & 1.81366  &   0 \\
\hline
$\delta \nu_2 \equiv \frac{|nu_2|}{6 \langle r^3 \rangle} $
& 0.37992 &  0.11412  & 0.04783 &    0 \\
\hline
$\delta\nu_3 \equiv \frac{|\nu_3}{8 \langle r^4 \rangle }$
&  0.35000 &  0.12937 & 0.05657 &    0  \\
\hline
\end{tabular}
\end{center}

\subsection{Application to the 1P case}

Now, let us turn to the $1P$ state.
The wavefunction for the $1P$ state is
$R_{1p}(r)Y_{1m}(\theta,\phi)$. For the
linear potential case, the differential equation for $R_{1p}(r)$ reads
\begin{eqnarray}
\Bigl[-(\frac{d^2}{dr^2}+{2\over r}\frac{d}{dr}) + \frac{2}{r^2}+r \Bigr]
 R_{1p}(r)=E~R_{1p}(r). \label{eq:PEQ1}
\end{eqnarray}
There is no analytical solution for $R_{1p}(r)$.
Although one can solve it numerically,  if we
use the variational method to find an optimal TWF
for the $1P$ state,  we would not have a convenient exact solutions
like Airy function for the $S$ state to compare with. In this case, only
the hypervirial relations can serve
as a powerful measure to judge the quality of TWF.

Now, we write the optimal TWF in a more general form 
\begin{eqnarray}
R^{(n)}_{twf}(r)=N_n r^n e^{-a r^b}  \ \ \ \ \  (n=1),  \label {eq:TWFn}
\end{eqnarray}
with the normalization constant $N_n$ being
\begin{eqnarray}
N_n=(2 a)^{\frac{3+2 n}{2 b}} \sqrt{{\frac{b}
{\Gamma(\frac{3+2 n}{b})}}}\ \ \ \ (n=1).  \label {eq:Non}
\end{eqnarray}
It is very easy to prove that this TWF also satisfies the general
virial theorem, namely $\nu_1=0$.
Because we have
\begin{eqnarray}
\frac{\partial R^{(n)}_{twf}(r)}{\partial a}
   = (\frac{3+2n}{2 a b}-r^b) R^{(n)}_{twf}(r) \label {eq:RN1}
\end{eqnarray}
and
\begin{eqnarray}
iWR_{twf}^{(n)}(r)= ( \frac{3}{2}+n-a b r^b ) R_{twf}^{(n)}(r),
\label{eq:RN2}
\end{eqnarray}
comparing (\ref {eq:RN1}) with (\ref {eq:RN2}), we can immediately reach
Eq.(\ref {eq:RAW}). Actually, $R_{twf}(r)$ is a special case of
TWF $R_{twf}^{(n)}(r)$ with $n=0$.

Taking $n=1$, one has the $1P$ state TWF
\begin{eqnarray}
R^{(1)}_{twf}(r)=N_1 r e^{-a r^b}. \label{eq:PR0}
\end{eqnarray}
We can  fix  the optimal values of the
variational parameter $a$ and the energy $E$ for $b=1, \frac{3}{2},
\frac{7}{4}$ and $2$, respectively,
according to the normal procedure, and the results are shown in Table 3.
The analytical expressions and their numerical results of
the deviation from the first and second hypervirial relations
$\nu_1, \nu_2 $ and $ \delta \nu_2$ are also shown in the table.

It is easy to see from Table 3 that

(a) $\nu_1$ is indeed zero, which is independent of the value of $b$. This is
consistent with our general discussion on the TWF
$R^{(n)}_{twf}(r)$;

(b) The optimal value of $a$ is different from the one for $1S$
state, and $\delta \nu_2$
can sensitively be used to select the best TWF. The result is 
$b=\frac{7}{4}$ again.
This conclusion is consistent with that achieved for the 1S state.
\vspace{0.5cm}

\centerline{ Table 3 }
\vspace{0.3cm}
{\footnotesize
Variational results with the single-parameter TWF Eq.(\ref{eq:PR0}) 
for the $1P$ state in the linear potential case. }

\begin{center}
\begin{tabular}{|c|c|c|c|c|}
\hline
  $b$ & 1&   $\frac{3}{2}$  & $\frac{7}{4}$ &  2\\
\hline
$a_0$&1.07722 & 0.47141 &  0.32323 &  0.22452 \\
\hline
E& 3.48119&3.36927 &  3.36135 & 3.36778 \\
\hline
$\nu_1 \equiv 3 \langle r \rangle-2 E$& 0 &0 & 0 & 0\\
\hline
$\nu_2 \equiv 5 \langle r \rangle+4 \langle \frac{1}{r} \rangle -
      4 E \langle r \rangle$&2.15443 &0.45059 & -0.02146& -0.39152\\
\hline
$\delta \nu_2 \equiv \frac{|\nu_2|}{5 \langle r^2 \rangle+4 \langle \frac{1}{r}
    \rangle}$&0.06250 &0.01467&0.00071 &0.01311\\
\hline
\end{tabular}
\end{center}

Instead, if we still employ a TWF in the form of
\begin{eqnarray}
R_{1p}(r)=c_0 r e^{-a_0 r^{\frac{7}{4}}},  \label {eq:1p}
\end{eqnarray}
where $a_0$ is taken to be the optimal value for the $1S$ state 
and $c_0$ is determined by the normalization condition. Then, 
there is no free parameter. Calculating $\langle H \rangle$,
one can straightforwardly obtain the energy 
     $$ E=3.6788.$$
The corresponding deviations from the first and second order
hypervirial relations are
\begin{eqnarray*}
\nu_1= -0.1003 \  \  \  \  \  \  \delta \nu_1=0.04676,
\end{eqnarray*}
and
\begin{eqnarray*}
\nu_2= -1.0472 \  \  \  \  \  \  \delta \nu_2=0.03760,
\end{eqnarray*}
respectively. This is because that the Hamiltonian for 
the $1P$ state is different 
from that for the $1S$ state, so that the adopted $a_0$ is not
the optimal value for the $1P$ state, and the $R_{1p}(r)$
given by (\ref {eq:1p}) does not satisfy the general virial theorem,
namely $\nu_1$ is not equal to zero. 
   
In comparison with the values of $\nu_2$ and $\delta\nu_2$
in Table 3 for $b=\frac{7}{4}$,
one can conclude that the TWF chosen in such a way is not an appropriate
one.

Now, we add one more variational parameter in and take TWF in the form of
\begin{eqnarray}
R_{1p}(r)=(c_0+c_1r) re^{-a_0 r^{\frac{7}{4}}}, \label {eq:1pa}
\end{eqnarray}
where $a_0$ still takes the value in Eq.(\ref{eq:1p}), $c_0$ is 
determined by the
normalization condition and $c_1$ is a new variational parameter.
By using variational method, we have
$$      E=3.36202 $$
and
$$\nu_1= -0.01177 \ \ \ \ \  \delta \nu_1= 0.00528,$$
$$\nu_2=-0.25348  \ \ \ \ \  \delta \nu_2=0.008526.$$
Evidently, the accuracy of TWF is much more improved.

\vspace{0.5cm}

\subsection{The other possible choice for the hypervirial operator}

One may choose different hypervirial operators $G$ instead of 
what we employed above.
Here, we give another example where the new hypervirial operator $G'$ is
in the form of
$$G^{\prime}=f(r) p_r^2.$$
With the Hamiltonian
$$H=p_r^2+V_{eff}(r)$$
where
 $$V_{eff}(r)=\frac{l(l+1)}{r^2}+V(r),$$
we have the commutator
\begin{equation}
[H,G^{\prime}]=f(r) \frac{d^2V_{eff}(r)}
     {dr^2}+2 i f(r) \frac{dV_{eff}(r)}{dr}~p_r
   -\frac{d^2f(r)}{dr^2} ~p_r^2-2~i \frac{df(r)}{dr}~p_r^3. \label {eq:cm3}
\end{equation}
 
Let us set $l=0$ and $V(r)=r$ for demonstration. Taking
$f(r)=r$, we obtain a new first order hypervirial relation
\begin{eqnarray}
\langle -i rp_r \rangle= \langle -ip_r^3 \rangle.  \label {eq:pp1}
\end{eqnarray}
Similar to above operation, we define
\begin{equation}
\nu_1^{\prime} \equiv  \langle -ip_r^3 \rangle - \langle -i rp_r \rangle
     \label {eq:pp2}
\end{equation}
and
\begin{equation}
\delta \nu_1^{\prime} \equiv \frac{|\nu_1^{\prime}|}{ \langle -i~p_r^3 \rangle}
     \label {eq:pp3}
\end{equation}
as a criterion to judge the quality of a TWF. However, considering
$$ p_r^2=H-r,$$
$$p_r r+r p_r=-i+2r p_r$$
and Eq.(\ref {eq:pp1}),  it is possible to give
an alternative expression for the criterion, which reads
\begin{equation}
\nu_1^{\prime\prime} \equiv E \langle -ip_r \rangle +1- 2 \langle -i r p_r
        \rangle.     \label {eq:pp4}
\end{equation}
Because for any eigenstate $|R_{eig}\rangle$ of the Hamiltonian
$H$ we have
\begin{equation}
H|R_{eig}\rangle=E|R_{eig}\rangle,
\end{equation}
$\nu_1^{\prime} $ is identical to $\nu_1^{\prime\prime}$ for the exact 
solution. For an approximate solution,
these two expressions would give different results. However, it is not
difficult to prove that for any function $R(r)=u(r)/r$ which 
satisfies the normalization condition
(\ref {eq:NU}) and the boundary condition (\ref {eq:UC}), we surely have
\begin{equation}
\langle -ip_r \rangle=0
\end{equation}
and
\begin{equation}
\langle -irp_r \rangle=\frac{1}{2}.
\end{equation}
Consequently, $\nu_1 ^{\prime\prime}$ vanishes trivially and cannot
be used as a criterion.
In general, $\nu_1^{\prime}$ is not zero. Therefore, $\nu_1^{\prime}$ and
$\delta \nu_1^{\prime}$ can be applied to judge the quality of a TWF.
We list the numerical result for the $1S$ state 
with the TWF parameters used in(\ref {eq:TWF}) in Table 4.

\vspace{0.5cm}

\centerline{ Table 4 }
\vspace{0.3cm}
{\footnotesize
Variational results with the single-parameter TWF (\ref {eq:TWF}) for 
the $1S$ state in the linear potential case.
The employed hypervirial operator is $G'$.}

\begin{center}
\begin{tabular}{|c|c|c|c|c|}
\hline
  $b$ & 1&   $\frac{3}{2}$  & $\frac{7}{4}$ &  2\\
\hline
$\nu_1^{\prime}$&  1 & 0.16667 & 0.018228 & -0.07559\\
\hline
$\delta \nu_1^{ \prime}$&0.6667 &0.25000 & 0.03517 & 0.17810\\
\hline
\end{tabular}
\end{center}

It is easy to see from  Table 4 that $\delta \nu'_1$ can indeed be used as
a criterion to judge the quality of the TWF and gives a consistent result
with those in Table 1, but it is less sensitive than $\delta\nu_2$
presented in Table 1.

\vspace{1cm}

\section{Conclusion and discussion}

\vspace{0.5cm}

There are a variety of approximation methods which can be employed to
solve the 
non-relativistic Schr\"{o}dinger
equation where exact analytical solutions cannot be obtained. The variational
method is one of the most frequently adopted approaches because of its
apparent advantages. As well known,
the obtained eigenenergies and wavefunctions
are closely related to the form of the variational trial wavefunction (TWF)
and the number of the variational parameters. What we concern is how close
to the real values the achieved results can be. This is a crucial problem which
should tell us the reliability of the results. In other words, we need a proper
way to judge the quality of the TWF, especially the single-parameter TWF,
because of its simplicity.

To answer this question, we can set some criteria which determine the quality
of the adopted TWF. In this work, we have carefully studied several calculable
quantities for this purpose. Some of them were suggested in literatures
and the rest of them are newly introduced by us. 
To obtain the idea of the criterion for judging the quality of TWF,
we use the simple potential form $V(r)=r$, which has exact 
solutions for $nS$ states, as an example.
The numerical results indicate that all the criteria are consistent, namely,
if one of them shows a sufficiently smaller deviation of the obtained result
from the exact solution, the others confirm the conclusion. However,
except the hypervirial relations, evaluating the quantities which stand
for the criteria of the quality of TWF needs the exact solutions of 
the Schr\"{o}dinger equation. Apparently, such solutions are generally 
not available. Only can the deviations from the hypervirial relations serve
as a practical
and powerful criterion for
the quality of TWF. Furthermore, in our special case with $V(r)=r$, we can
deduce a conclusion that the hypervirial relations can be a good 
criterion and is supported by the other criteria gained in the special 
case where the exact solutions for $nS$ states are available. This is also
valid for various potential forms. To consolidate our
confidence, we have studied the $1P$ state with $V(r)=r$, where no
exact solution exists. The results
are qualitatively and quantitatively consistent with what we expected.

Moreover, we use $G=r^Np_r$ as the hypervirial operator which is of
obvious physical significance. It should be specially noticed that
in some special or accidental cases, the first (maybe the first a few) 
hypervirial relations precisely hold. However, it is not due to the 
correctness of TWF, but the TWF's fall in the special
categories which are specified in Sec.III(D). In these cases, the
first a few hypervirial relations cannot be used as 
criteria for the quality of TWF, instead
one needs to invoke higher orders hypervirial relations or use other 
hypervirial operator, for example, $G'$ in Sec.III(E).
Thus, when one applies the hypervirial relations as criteria for
the quality of TWF, he has to study the form of the adopted TWF to
see if it falls into the categories listed in Sec.III(D).

Our conclusion affirms that the hypervirial relations hold 
precisely for the real eigenstates, but approximately for TWF. 
They can be used to determine the quality of the adopted TWF. 
The deviation of the hypervirial
relations from equality can serve as the most practical and powerful
criterion for choosing an appropriate TWF in any potential cases.

\vspace{1cm}

\acknowledgements

One of the authors (Y. B. Ding) would like to thank Prof. G. Prosperi for his
hospitality during his stay in the Department of Physics, University of
Milan. This work is partly supported by the National
Natural Science Foundation of China (NSFC), the Chinese Academy of Sciences
(CAS) and Istituto Nazionaale di Fisica
Nuclear of Italy (INFN)

\appendix

\section{The analytical solution of the Sch\"{o}dinger equation for $S$-states
in the linear  potential case}

In the framework of the non-relativistic potential model, with a
central potential $V(\vec r)=V(r)$,
the eigenstate energy $E$ and the corresponding radial wave function
$R(r)$ can  be obtained
by  solving the Schr\"{o}dinger equation
\begin{eqnarray}
H~R(r)\equiv
 \Bigl[\frac{p_r^2}{2\mu} + \frac{l(l+1)}{2\mu r^2}+ V(r) \Bigr]
 R(r)=E~R(r), \label{eq:SE1}
\end{eqnarray}
where $H$ is the Hamiltonian, $\mu$ the reduced mass, $p_r$ the radial
momentum operator and $l$ the quantum number of the angular momentum.
The radial wave function $R(r)$ satisfies the normalization condition:
\begin{equation}
\int^\infty_0~R^2(r)~r^2~dr=1\ .\label{eq:NR}
\end{equation}
Taking the linear potential $V(r)=r/r_0$ normalized as
$r_0=1$/GeV$^2$ (so we ignore $r_0$ in all later expressions),
$2\mu=1$ GeV and the
reduced radial wave function  $u(r)=r ~R(r)$  which satisfies the normalization
condition
\begin{equation}
\int^\infty_0~u^2(r)~dr=1, \label{eq:NU}
\end{equation}
and the boundary condition
\begin{equation}
u(0)=0 \ \ \ \ and \ \ \ u(\infty)=0.  \label{eq:UC}
\end{equation}
The equation for the $S$ state has a simple form:
\begin{equation}
\bigl(-\frac{d^2}{dr^2}+r\bigr)u(r)=Eu(r). \label{eq:SE2}
\end{equation}
Rewriting the equation as
\begin{equation}
\bigl(\frac{d^2}{dr^2}+(E-r)\bigr)u(r)=0, \label{eq:SE3}
\end{equation}
and comparing it with the Airy equation \cite{AIRY}
\begin{equation}
\bigl(\frac{d^2}{dz^2}-z \bigr) ~W(z)=0, \label{eq:SW}
\end{equation}
one easily obtains
\begin{equation}
u_{nS}(r)=N_n~Ai(r-E_n),  \ \ \ \ \   (n=1,2,3,\cdots \ \
        {\rm and} \ \ r\geq ~0)
\label{eq:U1}
\end{equation}
where $Ai(z)$ is the Airy function, $N_n$ the normalization constant 
and $E_n$ the eigenenergy of the $nS$ state in the linear potential
case, which is just the value of the negative $n$-th node of 
the Airy function $Ai(z)$. The numerical values of $E_n$ are 
$E_1=2.33810 \cdots$, $E_2=4.08794\cdots$, 
$E_3=5.52055\cdots$ etc., respectively. 

\section{The variational solution for the $S$-states in the linear
      potential case}

For the ground state, i.e. the $1S$ state, in the linear potential case
with $2\mu=1 GeV$, the nomarlized trial wave function (TWF) with a single 
variational parameter is taken to be
\begin{eqnarray}
     R_{twf}(r)=N~ e^{-a~r^b}, \label {eq:TWF}
\end{eqnarray}
with the normalization constant 
\begin{eqnarray}
    N=\Big[\frac{b~(2a)^{\frac{3}{b}}}
    {4 \pi \Gamma(\frac{3}{b})}\Big]^{\frac{1}{2}}.  \label {eq:NO}
\end{eqnarray}
In above equations, $a$ denotes the variational parameter
and $ b $ represents the model parameter which specifies the
type of the TWF. If one considers the linear potential only, 
the Hamiltonian for the $nS$ state can be written as 
\begin{eqnarray}
H=-\frac{d^2}{dr^2}-\frac{2}{r}~\frac{d}{dr} + r, \label{eq:SH1}
\end{eqnarray}
Solving the Sch\"{o}dinger equation in terms of the TWF  Eq.(\ref
{eq:TWF}), one obtains the analytical results for $a_0$ and $E$.
The optimal value of $a$ is
\begin{eqnarray}
a_0=\frac{1}{2} \Bigl( \frac{2\Gamma(\frac{4}{b})}{(1+b)\Gamma
(\frac{1}{b})}\Bigr)^{\frac{b}{3}},     \label {eq:A0}
\end{eqnarray}
and the corresponding minimum value of the energy reads
\begin{eqnarray}
E_{min}=\frac{3}{2\Gamma(\frac{3}{b})} \Bigl(
  \frac{(1+b)\Gamma(\frac{1}{b})\Gamma^2(\frac{4}{b})}{2}\Bigr)^
  {\frac{1}{3}}.     \label {eq:E0}
\end{eqnarray}

Because the behavior of the linear potential lies between
the Coulomb potential and the harmonic oscillator potential, we 
select four kinds of TWF with $b=1$ ( "hydrogen-like" wave function ) ,
$b=\frac{3}{2}$, $b=\frac{7}{4}$ and
$b=2$ ( Gaussian wave function ), respectively.                         
The resultant $a_0$ and $E_{min}$ for various values of $b$ are
shown in Table 1. It is clear that when $b=7/4$, one reaches the
most accurate solution. 

Taking $b=7/4$, one can construct a TWF for the $2S$ state. It reads
\begin{eqnarray}
R_{twf}^{2S}(r)=(c_0+c_1 r+c_2 r^2) e^{-a_0~r^{\frac{7}{4}}}, \label {eq:TWF2}
\end{eqnarray}
where $a_0=0.348957$, which is the optimal value determined for the $1S$ 
state, $c_0$ can be given by the normalization
condition, $c_1$ should be determined by the orthogonal condition to the
$1S$ state and $c_2$ is the unique variational parameter which can be
obtained by minimizing the corresponding average value of
the Hamiltonian $H$, $\langle R^{2S}_{twf}|H|R^{2S}_{twf}\rangle$.

Similarly, for the $3S$ state, we can take TWF as
\begin{eqnarray}
R_{twf}^{3S}(r)=(c_0+c_1 r+c2 r^2+c_3 r^3) e^{-a_0~r^{\frac{7}{4}}}.
\label {eq:TWF3}
\end{eqnarray}
The numerical values for the $2S$ and $3S$ states are also 
shown in Table 1.\\

\end{document}